Article

# Detection of Sulfur Dioxide by Broadband Cavity-Enhanced Absorption Spectroscopy (BBCEAS)

Ryan Thalman [1,*], Nitish Bhardwaj [2], Callum E. Flowerday [2] and Jaron C. Hansen [2]

1 Department of Chemistry, Snow College, Richfield, UT 84701, USA
2 Department of Chemistry and Biochemistry, Brigham Young University, Provo, UT 84602, USA; nitishb@byu.edu (N.B.); callumflowerday@gmail.com (C.E.F.); jhansen@chem.byu.edu (J.C.H.)
* Correspondence: ryan.thalman@snow.edu

**Abstract:** Sulfur dioxide ($SO_2$) is an important precursor for the formation of atmospheric sulfate aerosol and acid rain. We present an instrument using Broadband Cavity-Enhanced Absorption Spectroscopy (BBCEAS) for the measurement of $SO_2$ with a minimum limit of detection of 0.75 ppbv (3-$\sigma$) using the spectral range 305.5–312 nm and an averaging time of 5 min. The instrument consists of high-reflectivity mirrors (0.9985 at 310 nm) and a deep UV light source (Light Emitting Diode). The effective absorption path length of the instrument is 610 m with a 0.966 m base length. Published reference absorption cross sections were used to fit and retrieve the $SO_2$ concentrations and were compared to fluorescence standard measurements for $SO_2$. The comparison was well correlated, $R^2$ = 0.9998 with a correlation slope of 1.04. Interferences for fluorescence measurements were tested and the BBCEAS showed no interference, while ambient measurements responded similarly to standard measurement techniques.

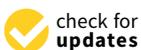

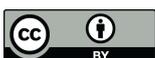

## 1. Introduction

Sulfur dioxide ($SO_2$) is a precursor to the formation of atmospheric sulfate aerosol and acid rain [1]. $SO_2$ is emitted naturally through volcanic eruption [2], oxidation of other atmospheric sulfur species [3], and is emitted anthropogenically from the oxidation of sulfur from the combustion of coal and oil [4,5]. $SO_2$ directly affects health through the respiratory system with elevated risks for high-risk groups [6].

Further oxidation of $SO_2$ can form sulfate ($SO_4^{2-}$) which in the form of sulfuric acid ($H_2SO_4$) contributes to acid rain but also contributes to particulate aerosol in the atmosphere [5]. Stratospheric injection of $SO_2$ by volcanoes and subsequent formation of stratospheric aerosol has been proven to have a short-term cooling effect on the global climate [7] and, therefore, is also being considered in some geoengineering scenarios as a possible technique to cool the global climate [8,9]. Even after decreased $SO_2$ emission by the United States and Europe, continued industrialization in other countries has seen an increase in global $SO_2$ emissions since 2000 [4].

There are several well-established measurement techniques for $SO_2$ that have been used in routine air-quality monitoring for decades, including UV fluorescence [10,11] and the pararosaniline wet chemistry technique, [12] which are the two Environmental Protection Agency (EPA) Federal Reference and Equivalent Methods [13]. Other techniques include photoacoustic spectroscopy [14], Cavity Ring-Down Spectroscopy [15], Long Path Differential Optical Absorption Spectroscopy (LP-DOAS) [16,17], Mass Spectrometry [18], and Multi-Axis Differential Optical Absorption Spectroscopy (MAX-DOAS) [19]. The LP- and MAX-DOAS techniques are not in situ measurements, but leverage the spectroscopic signature of $SO_2$ for quantification. The most common technique is UV fluorescence, with several different manufacturers selling instruments for monitoring. One such instrument, the 43i-Trace Level Enhanced from Thermo Electron Corp. (TECO, Franklin, MA, USA) has a detection limit of 0.2 ppbv for a 10 s average but can be as low as 0.05 ppbv for a 300 s, average with a precision of 1% of the measured concentration of 0.2 ppbv, based on the supplied manufacturer specifications. UV fluorescence uses pulsed UV light to excite the $SO_2$ molecules which then relax to re-emit light at a longer wavelength. The 43i instrument includes a hydrocarbon scrubber to remove most interfering hydrocarbons that also fluoresce when excited with UV light. Known interfering species for fluorescence technique include NO, m-xylene, and $H_2O$.

Broad Band Cavity Enhanced Spectroscopy (BBCEAS) leverages a high finesse optical cavity of a given wavelength to realize long path lengths, similar to LP- and MAX-DOAS but with in situ sampling. Related techniques often add the light source type in front of the acronym (Light Emitting Diode (LED, [20]), Interband Cascade Laser (ICL), or Optical Feedback (OF)). BBCEAS and related techniques have been used to measure species including: $NO_2$ [20–23], $NO_3$ [20,24], $H_2O$ [23], $O_3$ [23,25], glyoxal [22,23,26], methyl glyoxal [23], biacetyl [26], IO [23], $I_2$ [20,27], OIO [28], OClO [29], ClOOCl [30], BrO [31], HONO [32], HCHO [29,33], BTEX (benzene, toluene, ethyl benzene, and xylenes) compounds [34], and $O_4$ [20,35] in the UV and visible regions of the spectrum as well as other compounds in the near-IR and IR using related techniques such as Cavity Ring-down Spectroscopy (CRDS) [36], Integrated Cavity Output Spectroscopy (ICOS) [37], and Cavity Attenuated Phase Shift Spectroscopy (CAPS) [38]. $SO_2$ was recently measured using OF-CEAS at 4.035 μm with a detection limit of 130 ppbv [39] and CRDS in the UV with a detection limit of 3.5 ppbv in 10 s [15]. Previously $SO_2$ was measured by BBCEAS but in the range of 368–372 nm and at concentrations of 0.039–1% [40], as well as in the range of 250–280 nm as a calibration gas [34].

Spectroscopic measurements of $SO_2$ in the UV region are based on its highly structured absorption at wavelengths shorter than 320 nm. The structured absorption allows for independent quantification of $SO_2$ [41] from other gases that absorb in the same wavelength window including $NO_2$ [42], BrO [43], OClO [41], and many organic molecules with broad absorptions in the UV, acetone being just one example [44] (Figure 1). This work provides data for a BBCEAS instrument in the range of 305–312 nm using $SO_2$ as test molecule and preparing the way for further measurement possibilities of other UV absorbers.

The UV spectral region below 315 nm represents a relatively underexplored region for atmospheric detection of organic and other atmospherically relevant molecules by cavity-enhanced methods. In the past, this has been limited by both light source availability (cavity-ring down spectroscopy requires frequency doubling and a dye laser, LEDs were weak or not available at given wavelengths) and by poor mirror reflectivity. Washenfelder et al. [33] outlined the limitations of UV cavity-enhanced spectroscopy which include the lack of bright light sources as well as increasing mirror substrate and coating absorption and scattering losses that limit light throughput and mirror reflectivity. With the introduction of new, brighter LED light sources in the UV, this work explores the possibilities of utilizing the one portion of these UV wavelengths for detection of atmospherically relevant molecules.

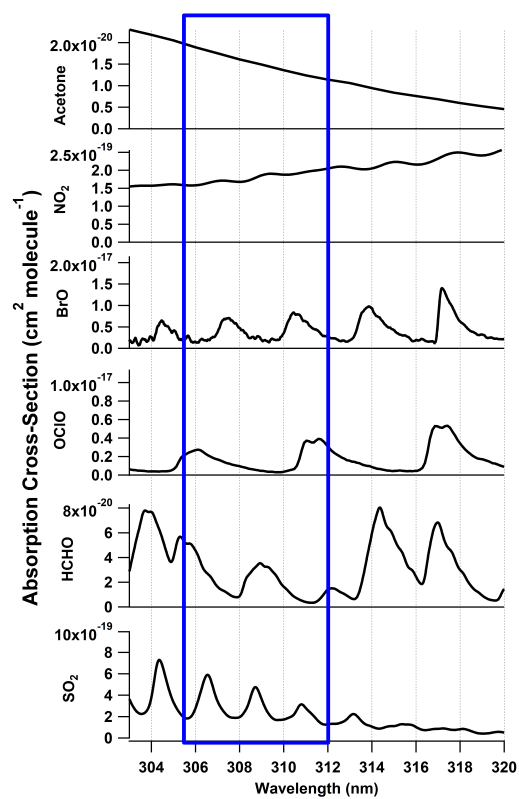

**Figure 1.** Absorption cross sections of species absorbing in the 300–320 nm range including, $SO_2$ [45], $NO_2$ [42], BrO [43], OClO [41], acetone [44], and HCHO [46]. The spectral fitting window for the $SO_2$ BBCEAS is shown in blue.

## 2. Materials and Methods

The $SO_2$ cavity instrument consists of an optical cavity mounted in a 3D-printed cage assembly sitting on top of an instrument control box. Figure 2 shows a schematic of the instrument including standard dilution and supply as well as gas control valves. The BBCEAS instrument consists of a light source (LED), collimating and focusing optics, the optical cavity, and an optical fiber leading to the detector (spectrograph). A UV LED with a center wavelength of 310 nm (Roithner–Lasertechnik GmbH, DUV310-SD353E) was attached to a printed circuit board with an output power of 50 mW collimated by a 25 mm f/1 UV lens. The LED was temperature controlled with a Peltier cooler to $14.0 \pm 0.2$ °C. The Peltier cooler consisted of a temperature controller (Omega) using a Type K thermocouple held on the front of the printed circuit board (PCB) by a 3D-printed brace holding the LED and Peltier cooler to the heat sink, and a small Peltier module mounted directly behind the LED (CUI Devices, Tualatin, OR, USA, CP30138, $15 \times 15 \times 3.8$ mm) (Figure S1 in the Supplementary Information). The manufacturer specifications state a nominal power output of 93% of initial power after 3000 h at 25 °C ambient temperature and 350 mA current. For 6 months of continuous use with the LED actively cooled and driven at 400–500 mA, no noticeable degradation of LED output was observed. The optical cavity consists of a pair of 2.5 cm diameter high reflectivity mirrors with a center wavelength of 310 nm, a stated maximum reflectivity of 99.9%, and a radius of curvature of 100 cm (Layertec GmbH). Filtered sample air enters and exits the cavity by the mirrors, utilizing the entire cavity length [47]. The cavity length was $96.6 \pm 0.1$ cm. The instrument including temperature control, valve control, data acquisition, and LED power supply uses <40 W of power (110 VAC).

The instrument temperature was measured by use of the on-board thermocouple which the LabJack U6 data acquisition system includes. The pressure was measured in the cavity airstream using a small pressure sensor (Honeywell, ASDXACX015PAAA5, 0–15 PSI) and the acquired voltages logged on the LabJack. The pressure sensor was inserted into the fitting near the mirror purge. The cavity flow was measured using a Honeywell 0–5 standard cubic centimeters per minute (sccm) flow meter (AWM5101VN) and the analog voltage output of the flow meters was logged using the LabJack.

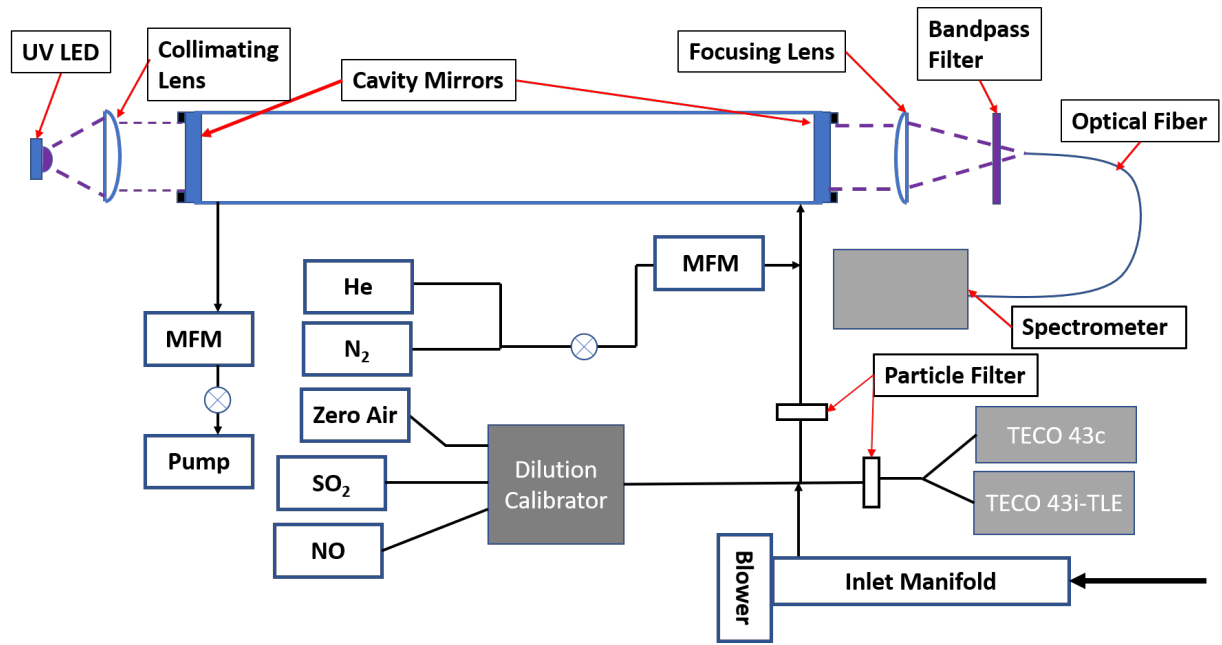

**Figure 2.** Schematic of BBCEAS cavity as set up for comparison with the $SO_2$ standard and ambient sampling. Flow is pulled into the system, and the total flow of the sample and the overflow are measured by mass flow meters (MFM). BBCEAS measurements are made in parallel with the Thermo Electron Corporation (TECO) 43 series instruments.

The optical cavity consists of a 0.75-inch outer diameter PFA Teflon tube placed between the mirrors. Light exiting the cavity is focused onto an optical fiber (Thorlabs, 6 × 200 µm round to linear bundle) by a 1-inch f/4 lens and filtered by a 12.5 mm-diameter bandpass filter (10 nm FWHM, 310 nm, Edmund Optics). The fiber is then directed to the slit of an Andor DU440-BV Spectrograph with a SR-303i CCD camera cooled by a Peltier cooler to −20 °C with a 1200 grooves/mm grating. The slit was set at 75 µm for a resolution of 0.25 nm FWHM, which provided a sharp well-defined line function. The fiber assembly only illuminated 100 rows of the 512 row detector, so a portion of the CCD was not read out for each scan in an effort to increase the signal-to-noise ratio. The CCD was set to an integration time of 0.2 s, with a readout time of 0.06 s. In total, 110 scans were coadded before saving, giving a minimum integration time of 30 s in these experiments. The optics are all mounted in an optical cage system constructed of pultruded carbon fiber tubes with the braces for the tubes made of 3-D printed parts consisting of Polylactic Acid (PLA) printed on an Ender3 (Creality) printer (see Figure 3). PLA was used for structural cage supports other than the mirror mounts since PLA is an easy material to print but does not provide an air-tight seal between layers. The 3D-printed parts were only used for structural support with stainless steel tubes inserted into the mirror mounts which sealed via an O-ring to the cavity mirrors. The Teflon tube was held in place between the two stainless steel tubes on each end using a bored-through pipe connection. The cavity was configured and operated without purge volumes, with the sample being pulled through a particle filter (Pall, 2 µm pore, 47 mm diameter) in a Teflon filter holder. The reflectivity of the optical cavity was measured using the differential Rayleigh scattering of He and $N_2$ gas according to the following equation [23]:

$$R(\lambda) = 1 - d_0 \frac{\left(\frac{I_{N_2}(\lambda)}{I_{He}(\lambda)}\right)\epsilon_{Ray}^{N_2}(\lambda) - \epsilon_{Ray}^{He}(\lambda)}{1 - \left(\frac{I_{N_2}(\lambda)}{I_{He}(\lambda)}\right)} \quad (1)$$

where $d_0$ is the cavity length (96.6 ± 0.1 cm), $\epsilon_{Ray}$ is the extinction due to Rayleigh scattering of the respective gases [48] and $I$ is the spectral intensity in the respective gas ($N_2$ or He). The measured reflectivity was found to be 99.85%, and the measured reflectivity, effective pathlength and example spectra for $N_2$ and He are given in Figure 4.

The measured concentrations were retrieved by nonlinear least-squares fitting of the cavity extinction as given by Fiedler et al. [49] and Washenfelder et al. [22]:

$$\epsilon(\lambda) = \left(\frac{1 - R(\lambda)}{d_0} + \epsilon_{\text{Rayleigh, Air}}(\lambda)\right)\left(\frac{I_0(\lambda) - I(\lambda)}{I(\lambda)}\right) \quad (2)$$

where the $\epsilon(\lambda)$ is the wavelength resolved extinction, $R(\lambda)$ is the mirror reflectivity, $d_0$ is the cavity base length, $\epsilon$ is the extinction due to Rayleigh scattering, $I_0(\lambda)$ is the reference spectrum, and $I(\lambda)$ is the measurement spectrum. The reference spectrum was obtained by overflowing the cavity with zero air (see Figure 4).

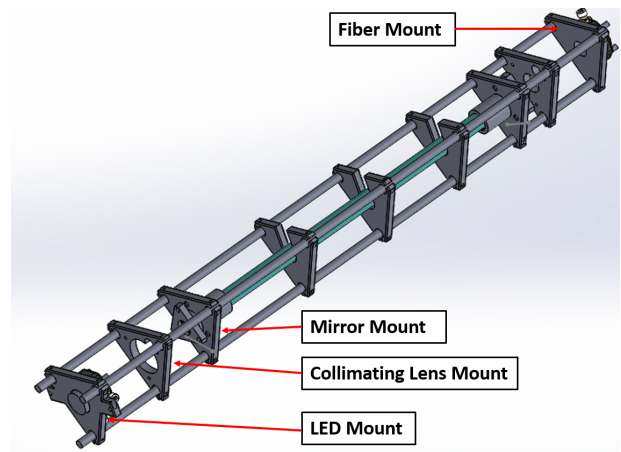

**Figure 3.** Mechanical drawing of the cage-mounting system for the BBCEAS. Cage plates are constructed of 3D-printed plastic parts with pultruded carbon tubes forming the optical cage.

The concentrations of the trace gases of interest were retrieved by nonlinear least square fitting in IGOR (Wavemetrics) by minimizing the error of the following equation with a 3rd-degree polynomial enabling a Differential Optical Absorption Spectroscopy retrieval [50]:

$$\epsilon(\lambda) = \sigma_{SO_2}(\lambda)[SO_2] + \sigma_{NO_2}(\lambda)[NO_2] + \text{polynomial}, \quad (3)$$

where $\sigma(\lambda)$ is the standard absorption cross section for the given gas and $[SO_2]$ is the retrieved concentration of $SO_2$ [41,45] and $[NO_2]$ [42]. The absorption cross sections were convolved to the instrument slit function using the convolution function in QDOAS [51]. Only $SO_2$ and $NO_2$ absorption was retrieved as the absorption cross sections of other gases are either too small or not in large-enough concentrations relative to the sensitivity of the instrument to be fitted. Cross sections of $SO_2$ and other possible absorbers are shown in Figure 1. Because of the fitted polynomial, the retrieval is only sensitive to the structured (differential) cross section and is insensitive to broad changes in the light source shape, aerosol scatter (if no filter was used), and other broadband absorbers (many organic compounds that interfere with fluorescence measurements, such as the acetone shown in Figure 1). Fitting was carried out from 305.5–312 nm using Equation (3) with a 3rd-order polynomial and the retrieved concentration was converted to a mixing ratio using the measured temperature and pressure.

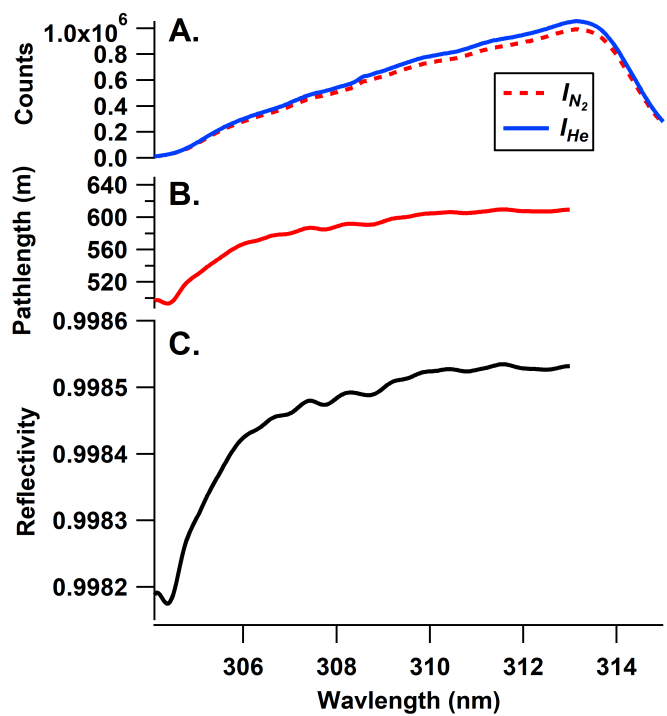

**Figure 4.** Panel (**A**): signal intensity in the presence of He and N$_2$ gas as used in Equation (1). Panel (**B**): effective pathlength (1/e) in meters. Panel (**C**): measured mirror reflectivity in the useable wavelength range.

*Comparison of SO$_2$ Measurements*

Initial testing of the BBCEAS instrument was carried out in comparison to an SO$_2$ standard cylinder (Airgas, 10.14 ppmv SO$_2$ in N$_2$, ±1.4%) diluted using a dilution calibrator (Environics, model 6103) which consists of two mass flow controllers (0–10 L per minute (lpm) and 0–50 sccm) diluting a small flow of the standard into a large flow of zero air providing a range of SO$_2$ concentrations from 0 to 170 ppbv. This range spans the normal operating range of the Thermo instruments as usually deployed (0–200 ppbv). The diluted standard was supplied to an inlet manifold which pulled air at a high flow (>20 lpm) and sampled into the BBCEAS through a 47 mm PTFE particle filter (Pall) using a pump at 1.0 lpm. The diluted standard was also sampled by TECO 43c and 43i-TLE SO$_2$ monitors to observe the response of the calibrator and ambient concentrations for the 43c. (See Figure 2. Data was logged internally on the 43i and via analog output (0–200 ppbv, 0–10 V).) Supplied concentrations were provided for a minimum of 10 min at each dilution setting to allow for the different settling times of the instruments and ensure enough overlap for averaging for correlation.

Interfering species were tested by introducing known interferences for the fluorescence instruments to the BBCEAS. NO was tested as an interfering species by injection using the same calibration setup as SO$_2$ with an NO standard of 20.42 ppmv NO (20.43 ppmv total NO$_x$ ± 2% in N$_2$). NO is the species reported by the manufacturers to have the largest interfering effect and is the species that is most likely to be encountered in ambient measurements at sufficient levels to have a significant influence on the measured SO$_2$ concentrations. Water vapor is considered an interference in stack sampling and *m*-xylene is a less commonly measured species compared to NO, with a lower reported interference response. Sampling and testing for NO proceeded in the following order: a sampling of ambient conditions; sampling of ambient conditions with standard addition of SO$_2$; sampling of SO$_2$ from the calibration cylinder in Zero Air at varying concentrations; followed by the return to ambient sampling. For xylenes as well as for the broadband absorber acetone, first, SO$_2$ was supplied from the dilution calibrator, after which this flow of diluted SO$_2$ standard was flowed separately through the head space of two flasks, one containing a mixture of xylenes, the other acetone (Sigma-Aldrich, Spec grade), to observe the change in the retrieved $SO_2$. Water was not tested as an interference, as the manufacturers state that this is only an issue with stack sampling which would be considered close to a condensing environment and can be adjusted for by adding inline dryers.

To evaluate the limit of detection, $N_2$ was continuously flowed through the cavity for 14 h with a 30 s minimum integration time. Spectra were then averaged over a given number of 30 s spectra to yield a maximum acquisition time of up to 400 min for both the spectra and the reference and evaluated with the Beer–Lambert law (absorption = $\ln(I_0/I)$) to assess the root mean square noise (RMS) over the fit window [50]. Pure photon counting noise follows the relationship $RMS = 1/\sqrt{N}$, where $N$ is the number of photons collected. To further assess spectral fitting performance with spectral averaging, the data series were also averaged to 1 min, 5 min, and 10 min intervals.

## 3. Results

### 3.1. Comparison to $SO_2$ Standard

The BBCEAS followed the response of the $SO_2$ concentration delivered by the dilution calibrator in a linear fashion. Figure 5 shows fitted extinctions at a range of different $SO_2$ concentrations and with different averaging times, highlighting unstructured residual features and good matching of the literature cross section to the data. Figure 6 shows the measured $SO_2$ concentrations. Several different measured conditions are highlighted in the figure, including ambient conditions, $SO_2$ standard addition to ambient sampling, and sampling of an $SO_2$ standard at a range of concentration levels. The correlation of the standard dilution from the calibrator with the BBCEAS retrieved concentrations yielded a slope of 1.04 ± 0.05, an offset of 0 ± 1 ppbv, and an $R^2$ value of 0.9998 (Figure 7). The absence of any structure in the residuals suggests no systematic error in the fitting routine and means that longer integration times and more acquired photons will lower the detection limit as expected from photon-shot noise. The Rufus et al. [45] cross section was used for fitting because the fit residual was improved by 20% at higher concentrations over the use of the Bogumil et al. [41] cross section, likely due to the fact that the Bogumil cross section is a lower resolution than our current instrument. For spectral-fitting purposes, the reference spectra were averaged from 10 min of zero-air spectra. The minimum fit residual for the 30 s average is $1.6 \times 10^{-8}$ cm$^{-1}$. The variability of the retrieved concentration at each concentration level indicated a limit of detection of 2.6 ppbv (3-$\sigma$) for a 30 s acquisition.

Data under ambient conditions showed that the two instruments followed each other within the operational parameters. Most of the ambient data exhibited no measured $SO_2$ as shown in the ambient portion of Figure 4 as well as the longer time period shown in Figure S2 in the Supplementary Materials as well as the correlation of the BBCEAS relative to the 43i-TLE for the same period (Figure S3).

### 3.2. Interferences

The fluorescence-based detection instruments reported measured $SO_2$ from NO injected into the sampling line, while the BBCEAS did not measure any NO when 4 ppmv of NO was injected (see Figure S4 in the Supplementary Information). The 43i registered a measured $SO_2$ concentration of 85 ppbv, giving a response of 0.085 ppbv $SO_2$ for every 1 ppbv of NO. This is a likely explanation for some of the baseline drift for the TECO 43i observed under clean conditions for $SO_2$, but with moderate $NO_x$. For xylenes (~1 ppmv) and acetone (~20 ppmv), no change in the measured $SO_2$ was observed for BBCEAS (Figure 8).

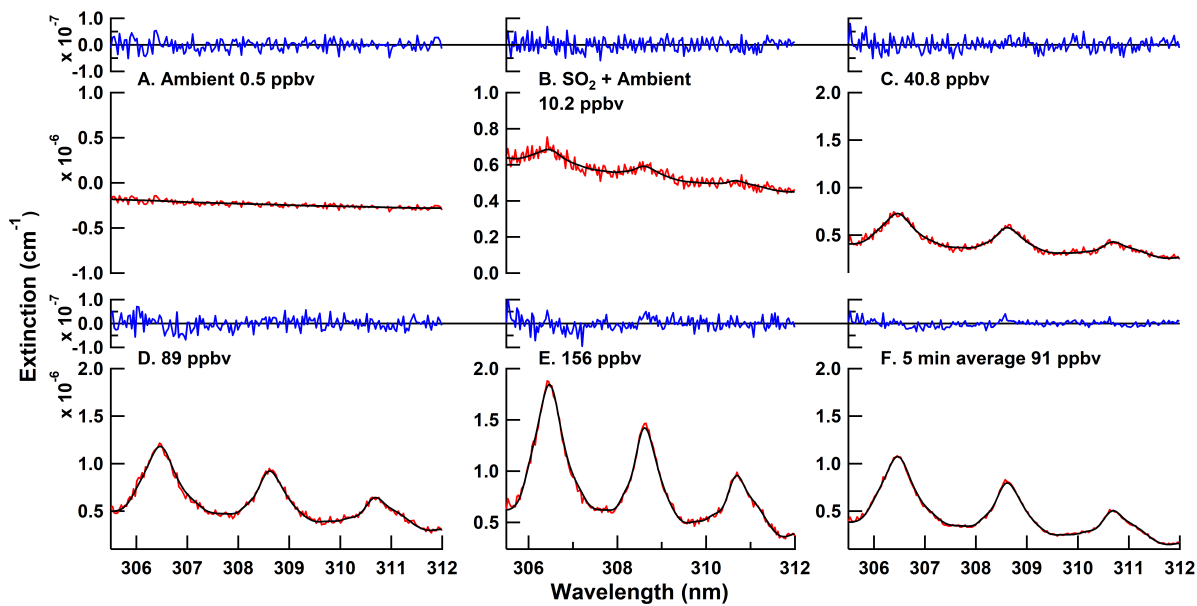

**Figure 5.** Fits of Equation (3) (black) relative to the measured extinction (red) in the lower portion of each panel. The difference of the black and red is shown on the upper axis for each panel in blue.

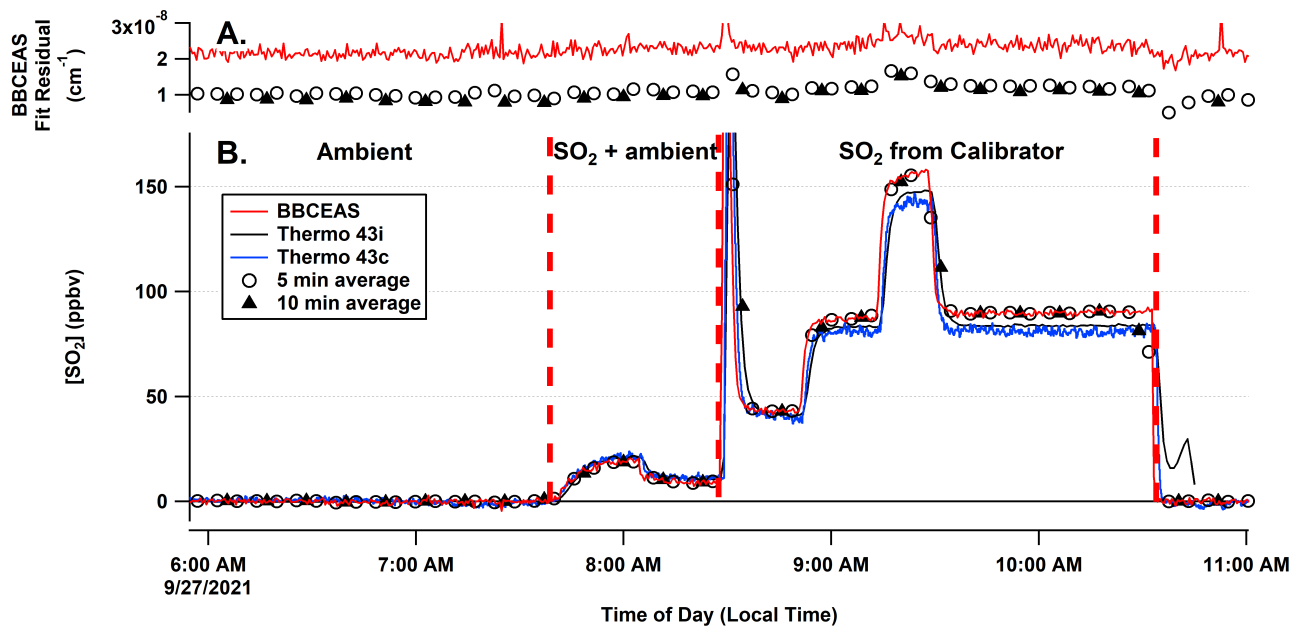

**Figure 6.** Time series of retrieved $SO_2$ concentrations. Panel (**A**) shows the 1-$\sigma$ standard deviation of the fit residual for the 30 s, 5 min, and 10 min data. Panel (**B**) shows the measured $SO_2$ from the three instruments under ambient, $SO_2$ + ambient, and calibration conditions as well as the time traces for the 5 and 10 min averaged data. Vertical red dashed lines separate the different conditions. The jump in the 43i signal at the end of the experiment is due to a flow connection change to that instrument.

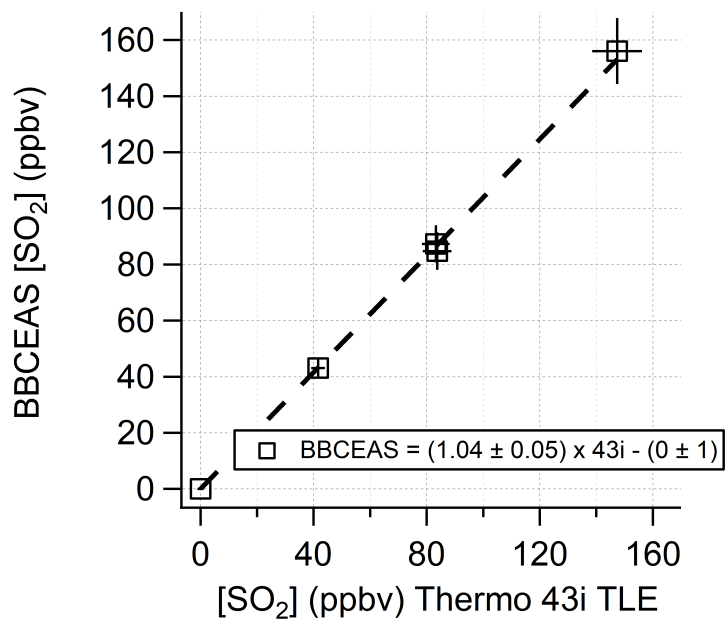

**Figure 7.** Correlation of BBCEAS (boxes) measured $SO_2$ with respect to that measured by the TECO 43i-TLE. The linear fit equation and uncertainties are included in the graph.

*3.3. Signal-Averaging Effect on Precision and Accuracy*

Signal-to-noise evaluation was carried out on spectra of $N_2$ with the Andor spectrometer using several hours of $N_2$ data. The data show no plateau for up to 20 min of integration time and a minimum RMS photon shot noise of $8.7 \times 10^{-5}$ (see Figure 9). Signal averaging yields 3-$\sigma$ detection limits in the fitted spectra of 2.6, 2.25, 0.75, and 0.48 ppbv for integration times of 30 s, 1 min, 5 min, and 10 min, respectively, as derived from the standard deviation of the measurement of the baseline for retrieved concentrations [26]. The overall uncertainty of the instrument measurement is limited by the fit RMS at low concentrations and by the cross section uncertainty (5% 1-$\sigma$) at higher concentrations [45]. The other contributing uncertainties are the measurement of the spectra (<1% based on the amount of signal acquired), the measurement of the pressure (5%), the measurement of the cavity length (<1%), and the mirror reflectivity (<2%, including the Rayleigh scattering cross section uncertainty [48]). The calculated extinction has an uncertainty of 2%, which, when combined with the absorption cross section uncertainty, gives an overall uncertainty of 5.4%. The values for the detection limits of several atmospherically relevant species as they can be extrapolated from the 5 min detection limit for $SO_2$ are shown in Table 1.

**Table 1.** Estimated limits of detection for a 5-min sampling time of atmospherically relevant species that absorb in the same wavelength range as $SO_2$.

| Species | $\sigma'$ * ($cm^3$ $molecule^{-1}$) | LOD (ppbv) | $\sigma$ Reference |
|---|---|---|---|
| $SO_2$ | $3.97 \times 10^{-19}$ | 0.75 | Rufus et al. [45] |
| $NO_2$ | $2.2 \times 10^{-20}$ | 13.5 | Vandaele et al. [42] |
| HCHO | $2.73 \times 10^{-20}$ | 10.9 | Meller and Moortgat [46] |
| OClO ** | $3.6 \times 10^{-18}$ | 0.09 | Dong et al. [29] |
| BrO ** | $6.6 \times 10^{-18}$ | 0.05 | Wilmouth et al. [43] |
| ClO ** | $3.7 \times 10^{-19}$ | 1.2 | Sander and Friedl [52] |

* The differential cross section is taken as the maximum peak-to-peak cross section in the 306–312 nm at the instrument resolution. ** Short lived species are not detectable with an inlet system.

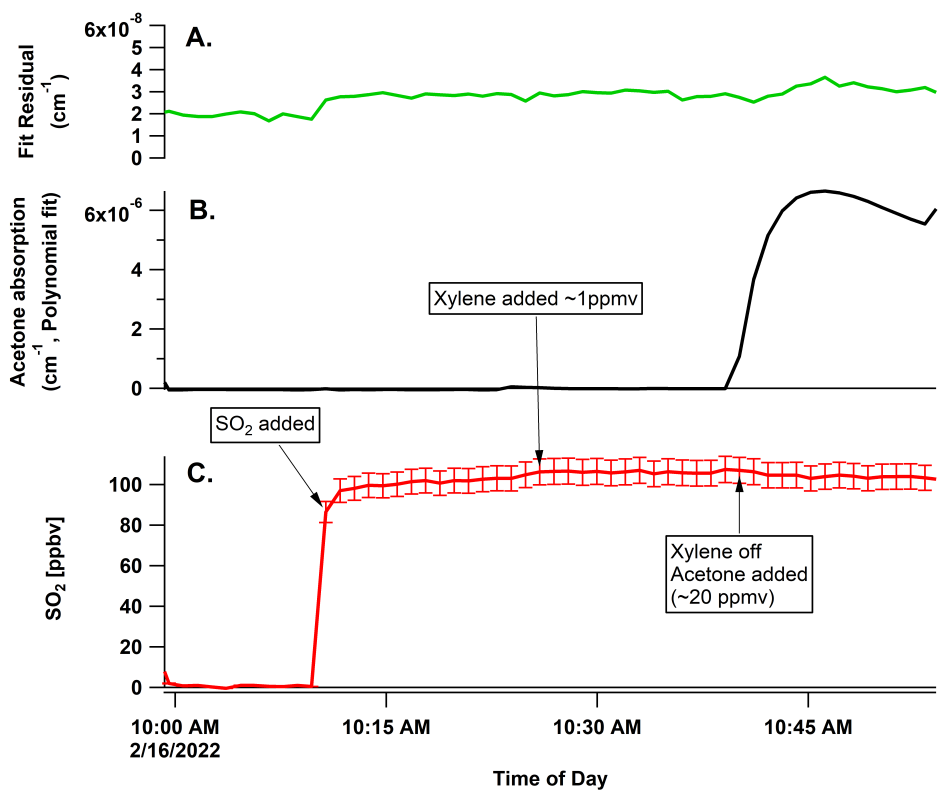

**Figure 8.** Panel (**A**)—fit residual RMS for interference of xylenes and acetone; panel (**B**)—fitted polynomial at 308 nm for xylene and acetone interference, showing fitted acetone absorption that is accomodated in the fit by the polynomial; panel (**C**)—retrieved concentrations of $SO_2$ in the presence of xylenes and then acetone.

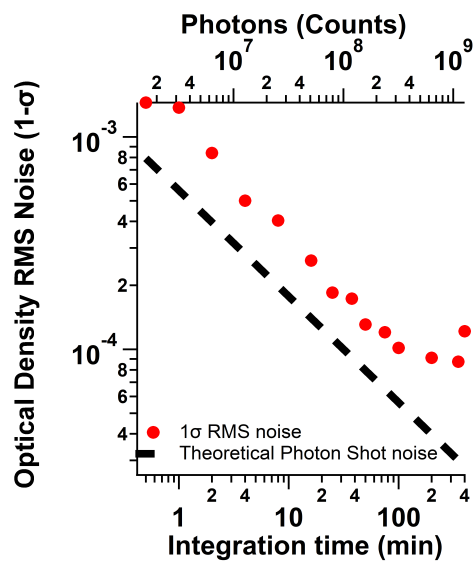

**Figure 9.** Signal-to-noise evaluation for the spectrometer evaluated as the 1-$\sigma$ RMS noise. The RMS noise levels off at longer integration times.

*3.4. Performance of 3D-Printed Cage System*

The 3D-printed cage system held up well under the movement of the instrument between locations including car trips. Weaknesses in the design include metal screws in plastic threaded holes and flexibility of the parts if exposed to excess heat. The cavity plates tended to crack if the attachment to the carbon tubes was tightened too much and excess heat (PLA deforms at 60 °C) from the LED cooling assembly once melted the cage plate holding the LED in the cage. While these are perhaps barriers for commercialization, the replacement parts were easily reprinted (<USD 1 each in materials and 6–12 h of printing) and replaced for the defective parts. Acrylic styrene-acrylonitrile (ASA) printed parts smoothed with acetone vapor were attempted to be used for the mirror mounts, but an airtight seal proved difficult to achieve, leading to the insertion of stainless-steel tubes.

## 4. Discussion

The BBCEAS instrument as currently constructed provides a complementary technique for the measurement of $SO_2$ with similar limits of detection and linearity over a wide range of $SO_2$ concentrations, comparable to what common commercial instruments for ambient monitoring can provide. In the current configuration, the 3-$\sigma$ detection limit is 2.6 ppbv for a 30 s integration time and 0.75 ppbv for a 5 min integration time. The instrument could be further optimized in terms of its light throughput and efficiency. Due to the broadened nature of the absorption lines of $SO_2$ (~1 nm FWHM), the instrument resolution of the Andor spectrograph (0.26 nm) was unnecessarily high. The ideal line width and grating combination would be 0.5 nm with a grating of 600 grooves/mm to maximize light throughput while maintaining a large enough differential absorption cross section for spectral fitting. This improvement in signal-to-noise would further improve the minimum detection limit and time response for trace level detection of $SO_2$ in the presence of other structured absorbers.

The instrument is calibrated with pure gases of $N_2$ and helium, removing the need for standards to be kept in the field for calibration, which is common practice for standard fluorescence techniques. Known interferences from NO, m-xylene (represented by a mixture of xylenes in this work), and $H_2O$ in the instruments utilizing fluorescence detection are avoided using the BBCEAS method as demonstrated for NO and for xylenes. Both of these compounds absorb light further into the UV (230–280 nm) and fluoresce similar to $SO_2$. Acetone is a broadband absorber which changes the total extinction inside the cavity. While the BBCEAS instrument is insensitive to retrieving concentrations of broadband absorbers (or scattering from aerosol), the instrument was still able to retrieve the $SO_2$ concentration in the presence of the acetone, demonstrating the insensitivity to broadband extinction processes as shown previously [23]. Additionally, it should be noted that early cavity enhanced spectroscopy works [49] omitted the scattering term in the extinction calculation (Equation (2)) for mirrors of lower reflectivity. Even at this low of mirror reflectivity, this yields a 4% error in the retrieved trace gas concentrations and should be included as first noted by Washenfelder et al. [22].

The cavity and spectrometer combination demonstrated here allows for signal averaging up to several hours of data with improved limits of detection. This has been previously demonstrated for cavity and DOAS fitting of spectra when the instrument behaves as a white-noise sensor, as has been demonstrated in this work [53]. The 3D-printed cage performed well and can be utilized for structural construction of optical cavities at a greatly reduced cost, or for researchers without access to machining and precision design support. While there are weaknesses to these construction techniques in terms of the use of plastic for cage supports, the plastic mounts still have a cost and weight advantage for accessibility of design, construction, and applications that require less weight.

In comparison to fluorescence techniques (similar to the Thermo instruments in this paper) BBCEAS is a direct absorption technique and therefore does not suffer from interference from cofluorescing species as shown in the intercomparison tests. Other absorbance measurements (LP or MAX-DOAS) depend on long path lengths in the atmosphere to and do not represent a point measurement. While both of these techniques can utilize wider fitting windows, thereby leveraging more channels to minimize cross-talk between absorbing species and decrease detection limits, these wider-fit windows also come with other challenges such as instrument straylight, scattered sunlight from long open paths (LP-DOAS) and possible interferences from the solar spectrum (MAX-DOAS, e.g., Fraun-hofer lines, Raman scattering). MAX-DOAS SO$_2$ measurements also fight the dropping intensity in ambient scattered light as the SO$_2$ absorption bands gets stronger at shorter wavelengths. Other point measurement techniques such as mass spectrometry, while very sensitive, with high time responses, require more complicated setup, high vacuum, and a large power consumption [18]. A BBCEAS instrument for SO$_2$ measurements has an advantage when applied to ambient point measurements, balancing size, power consumption, ease of calibration, and lack of interfering species.

## 5. Conclusions

This demonstration of workable BBCEAS measurements further into the UV spectral range with lower reflectivity mirrors allows for the measurement of a number of molecules of interest by BBCEAS in the UV and visible-light ranges. This work demonstrates a BBCEAS measurement of SO$_2$ with a 5 min detection limit of 0.75 ppbv, low enough for ambient air quality measurements. The BBCEAS also simplifies calibration with inert gases instead of traceable standards and is free from interfering species. Continued improvement of higher-powered UV LEDs provided enough light to access detection limit ranges of atmospheric importance (for SO$_2$ 0.5–200 ppbv) [33]. Other short-lived species may also be detectable by utilizing open-path detection schemes with longer cavity lengths (BrO, OClO, OH radical)[54]. Future development of the BBCEAS instrument could be made to lower the power requirements enough to allow the instrument to be mounted on a mobile platform such as an Unmanned Aerial System (UAS) for SO$_2$ source identification for large emitters.

**Supplementary Materials:** The following supporting information can be downloaded at: https://www.mdpi.com/article/10.3390/s22072626/s1, Figure S1: LED cooling mount in the cage system.; Figure S2: Ambient measurements of SO$_2$; Figure S3: Correlation of ambient data from Figure S3 for BBCEAS measured [SO$_2$] relative to the 43i; Figure S4: Response of BBCEAS vs. fluorescence detection in the presence of NO.

**Author Contributions:** Conceptualization, R.T. and J.C.H.; construction, data collection and analysis, R.T., N.B., C.E.F. and J.C.H.; writing, review, and editing, R.T. and J.C.H.; editing, N.B. and C.E.F. All authors have read and agreed to the published version of the manuscript.

**Funding:** This work was supported by the Utah NASA Space Grant Consortium's Faculty Research Seed Funding Award (RT), from the Department of Chemistry and Biochemistry's (BYU) Undergraduate Research Award program, and by the National Science Foundation (grant no. AGS-2114655).

**Institutional Review Board Statement:** Not applicable.

**Informed Consent Statement:** Not applicable.

**Data Availability Statement:** The dataset related to this article is hosted at ScholarsArchive managed by Brigham Young University (https://scholarsarchive.byu.edu/) (accessed on 20 February 2022).

**Conflicts of Interest:** The authors declare no conflict of interest.

## Abbreviations

The following abbreviations are used in this manuscript:

| | |
|---|---|
| BBCEAS | Broadband Cavity-Enhanced Absorption Spectroscopy |
| DOAS | Differential Optical Absorption Spectroscospy |
| PLA | Poly-lactic Acid |
| slpm | standard liters per minute |
| sccm | standard cubic centimeters per minute |
| ASA | Acrylic styrene-acrylonitrile |

**References**

1. Schwartz, S.E. Both Sides Now. *Ann. N. Y. Acad. Sci.* **1987**, *502*, 83–144. [CrossRef]
2. Holasek, R.E.; Self, S.; Woods, A.W. Satellite observations and interpretation of the 1991 Mount Pinatubo eruption plumes. *J. Geophys. Res. Solid Earth* **1996**, *101*, 27635–27655. [CrossRef]
3. Logan, J.A.; McElroy, M.B.; Wofsy, S.C.; Prather, M.J. Oxidation of $CS_2$ and COS: Sources for atmospheric $SO_2$. *Nature* **1979**, *281*, 185–188. [CrossRef]
4. Smith, S.J.; van Aardenne, J.; Klimont, Z.; Andres, R.J.; Volke, A.; Delgado Arias, S. Anthropogenic sulfur dioxide emissions: 1850–2005. *Atmos. Chem. Phys.* **2011**, *11*, 1101–1116. [CrossRef]
5. Hidy, G.M.; Blanchard, C. The changing face of lower tropospheric sulfur oxides in the United States. *Elem. Sci. Anthr.* **2016**, *4*, 000138. [CrossRef]
6. EPA. *Integrated Science Assessment for Sulfur Oxides-Health Criteria*; Technical Report EPA-HQ-ORD-2013-0357; U.S. Environmental Protection Agency: Washington, DC, USA, 2017.
7. Bluth, G.J.S.; Rose, W.I.; Sprod, I.E.; Krueger, A.J. Stratospheric Loading of Sulfur From Explosive Volcanic Eruptions. *J. Geol.* **1997**, *105*, 671–684. [CrossRef]
8. Rasch, P.J.; Tilmes, S.; Turco, R.P.; Robock, A.; Oman, L.; Chen, C.C.J.; Stenchikov, G.L.; Garcia, R.R. An Overview of Geo-engineering of Climate Using Stratospheric Sulphate Aerosols. *Philos. Trans. Math. Phys. Eng. Sci.* **2008**, *366*, 4007–4037. [CrossRef]
9. Visioni, D.; Pitari, G.; Aquila, V. Sulfate geoengineering: A review of the factors controlling the needed injection of sulfur dioxide. *Atmos. Chem. Phys.* **2017**, *17*, 3879–3889. [CrossRef]
10. Parrish, D.D.; Fehsenfeld, F.C. Methods for gas-phase measurements of ozone, ozone precursors and aerosol precursors. *Atmos. Environ.* **2000**, *34*, 1921–1957. [CrossRef]
11. Stecher, H.A., III; Luther, G.W., III; MacTaggart, D.L.; Farwell, S.O.; Crosley, D.R.; Dorko, W.D.; Goldan, P.D.; Beltz, N.; Krischke, U.; Luke, W.T.; et al. Results of the Gas-Phase Sulfur Intercomparison Experiment (GASIE): Overview of experimental setup, results and general conclusions. *J. Geophys. Res. Atmos.* **1997**, *102*, 16219–16236. [CrossRef]
12. West, P.W.; Gaeke, G.C. Fixation of Sulfur Dioxide as Disulfitomercurate (II) and Subsequent Colorimetric Estimation. *Anal. Chem.* **1956**, *28*, 1816–1819. [CrossRef]
13. Gilliam, J.; Hall, E. *Reference and Equivalent Methods Used to Measure National Ambient Air Quality Standards (NAAQS) Criteria Air Pollutants*; Technical Report EPA/600/R-16/139; U.S. Environmental Protection Agency: Washington, DC, USA, 2016; Volume I.
14. Yin, X.; Wu, H.; Dong, L.; Li, B.; Ma, W.; Zhang, L.; Yin, W.; Xiao, L.; Jia, S.; Tittel, F.K. ppb-Level $SO_2$ Photoacoustic Sensors with a Suppressed Absorption–Desorption Effect by Using a 7.41 µm External-Cavity Quantum Cascade Laser. *ACS Sens.* **2020**, *5*, 549–556. [CrossRef] [PubMed]
15. Medina, D.S.; Liu, Y.; Wang, L.; Zhang, J. Detection of Sulfur Dioxide by Cavity Ring-Down Spectroscopy. *Environ. Sci. Technol.* **2011**, *45*, 1926–1931. [CrossRef] [PubMed]
16. Stutz, J.; Platt, U. Improving long-path differential optical absorption spectroscopy with a quartz-fiber mode mixer. *Appl. Opt.* **1997**, *36*, 1105–1115. [CrossRef]
17. Lee, J.; Kim, K.H.; Kim, Y.J.; Lee, J. Application of a long-path differential optical absorption spectrometer (LP-DOAS) on the measurements of $NO_2$, $SO_2$, $O_3$, and $HNO_2$ in Gwangju, Korea. *J. Environ. Manag.* **2008**, *86*, 750–759. [CrossRef]
18. Speidel, M.; Nau, R.; Arnold, F.; Schlager, H.; Stohl, A. Sulfur dioxide measurements in the lower, middle and upper troposphere: Deployment of an aircraft-based chemical ionization mass spectrometer with permanent in-flight calibration. *Atmos. Environ.* **2007**, *41*, 2427–2437. [CrossRef]
19. Cheng, Y.; Wang, S.; Zhu, J.; Guo, Y.; Zhang, R.; Liu, Y.; Zhang, Y.; Yu, Q.; Ma, W.; Zhou, B. Surveillance of $SO_2$ and $NO_2$ from ship emissions by MAX-DOAS measurements and the implications regarding fuel sulfur content compliance. *Atmos. Chem. Phys.* **2019**, *19*, 13611–13626. [CrossRef]
20. Ball, S.M.; Langridge, J.M.; Jones, R.L. Broadband cavity enhanced absorption spectroscopy using light emitting diodes. *Chem. Phys. Lett.* **2004**, *398*, 68–74. [CrossRef]
21. Langridge, J.M.; Ball, S.M.; Jones, R.L. A compact broadband cavity enhanced absorption spectrometer for detection of atmospheric $NO_2$ using light emitting diodes. *Analyst* **2006**, *131*, 916–922. [CrossRef]
22. Washenfelder, R.A.; Langford, A.O.; Fuchs, H.; Brown, S.S. Measurement of glyoxal using an incoherent broadband cavity enhanced absorption spectrometer. *Atmos. Chem. Phys.* **2008**, *8*, 7779–7793. [CrossRef]
23. Thalman, R.; Volkamer, R. Inherent calibration of a blue LED-CE-DOAS instrument to measure iodine oxide, glyoxal, methyl glyoxal, nitrogen dioxide, water vapour and aerosol extinction in open cavity mode. *Atmos. Meas. Tech.* **2010**, *3*, 1797–1814. [CrossRef]
24. Venables, D.S.; Gherman, T.; Orphal, J.; Wenger, J.C.; Ruth, A.A. High Sensitivity in Situ Monitoring of $NO_3$ in an Atmospheric Simulation Chamber Using Incoherent Broadband Cavity-Enhanced Absorption Spectroscopy. *Environ. Sci. Technol.* **2006**, *40*, 6758–6763. [CrossRef] [PubMed]
25. Axson, J.L.; Washenfelder, R.A.; Kahan, T.F.; Young, C.J.; Vaida, V.; Brown, S.S. Absolute ozone absorption cross section in the Huggins Chappuis minimum (350–470 nm) at 296 K. *Atmos. Chem. Phys.* **2011**, *11*, 11581–11590. [CrossRef]

26. Thalman, R.; Baeza-Romero, M.T.; Ball, S.M.; Borrás, E.; Daniels, M.J.S.; Goodall, I.C.A.; Henry, S.B.; Karl, T.; Keutsch, F.N.; Kim, S.; et al. Instrument intercomparison of glyoxal, methyl glyoxal and $NO_2$ under simulated atmospheric conditions. *Atmos. Meas. Tech.* **2015**, *8*, 1835–1862. [CrossRef]
27. Dixneuf, S.; Ruth, A.A.; Vaughan, S.; Varma, R.M.; Orphal, J. The time dependence of molecular iodine emission from Laminaria digitata. *Atmos. Chem. Phys.* **2009**, *9*, 823–829. [CrossRef]
28. Vaughan, S.; Gherman, T.; Ruth, A.A.; Orphal, J. Incoherent broad-band cavity-enhanced absorption spectroscopy of the marine boundary layer species I2, IO and OIO. *Phys. Chem. Chem. Phys.* **2008**, *10*, 4471–4477. [CrossRef]
29. Dong, M.; Zhao, W.; Huang, M.; Chen, W.; Hu, C.; Gu, X.; Pei, S.; Huang, W.; Zhang, W. Near-ultraviolet Incoherent Broadband Cavity Enhanced Absorption Spectroscopy for OClO and CH2O in Cl-initiated Photooxidation Experiment. *Chin. J. Chem. Phys.* **2013**, *26*, 133–139. [CrossRef]
30. Young, I.A.K.; Jones, R.L.; Pope, F.D. The UV and visible spectra of chlorine peroxide: Constraining the atmospheric photolysis rate. *Geophys. Res. Lett.* **2014**, *41*, 1781–1788. [CrossRef]
31. Hoch, D.J.; Buxmann, J.; Sihler, H.; Pöhler, D.; Zetzsch, C.; Platt, U. An instrument for measurements of BrO with LED-based Cavity-Enhanced Differential Optical Absorption Spectroscopy. *Atmos. Meas. Tech.* **2014**, *7*, 199–214. [CrossRef]
32. Gherman, T.; Venables, D.S.; Vaughan, S.; Orphal, J.; Ruth, A.A. Incoherent Broadband Cavity-Enhanced Absorption Spectroscopy in the near-Ultraviolet: Application to HONO and $NO_2$. *Environ. Sci. Technol.* **2008**, *42*, 890–895. [CrossRef]
33. Washenfelder, R.A.; Attwood, A.R.; Flores, J.M.; Zarzana, K.J.; Rudich, Y.; Brown, S.S. Broadband cavity-enhanced absorption spectroscopy in the ultraviolet spectral region for measurements of nitrogen dioxide and formaldehyde. *Atmos. Meas. Tech.* **2016**, *9*, 41–52. [CrossRef]
34. Wang, M.; Varma, R.; Venables, D.S.; Zhou, W.; Chen, J. A Demonstration of Broadband Cavity-Enhanced Absorption Spectroscopy at Deep-Ultraviolet Wavelengths: Application to Sensitive Real-Time Detection of the Aromatic Pollutants Benzene, Toluene, and Xylene. *Anal. Chem.* **2022**, *94*, 4286–4293. [CrossRef] [PubMed]
35. Thalman, R.; Volkamer, R. Temperature dependent absorption cross-sections of $O_2$–$O_2$ collision pairs between 340 and 630 nm and at atmospherically relevant pressure. *Phys. Chem. Chem. Phys.* **2013**, *15*, 15371–15381. [CrossRef]
36. Brown, S.S. Absorption Spectroscopy in High-Finesse Cavities for Atmospheric Studies. *Chem. Rev.* **2003**, *103*, 5219–5238. [CrossRef] [PubMed]
37. O'Keefe, A.; Scherer, J.J.; Paul, J.B. CW Integrated cavity output spectroscopy. *Chem. Phys. Lett.* **1999**, *307*, 343–349. [CrossRef]
38. Kebabian, P.L.; Herndon, S.C.; Freedman, A. Detection of Nitrogen Dioxide by Cavity Attenuated Phase Shift Spectroscopy. *Anal. Chem.* **2005**, *77*, 724–728. [CrossRef]
39. Richard, L.; Ventrillard, I.; Chau, G.; Jaulin, K.; Kerstel, E.; Romanini, D. Optical-feedback cavity-enhanced absorption spectroscopy with an interband cascade laser: Application to $SO_2$ trace analysis. *Appl. Phys. B* **2016**, *122*, 247. [CrossRef]
40. Chen, J.; Venables, D.S. A broadband optical cavity spectrometer for measuring weak near-ultraviolet absorption spectra of gases. *Atmos. Meas. Tech.* **2011**, *4*, 425–436. [CrossRef]
41. Bogumil, K.; Orphal, J.; Homann, T.; Voigt, S.; Spietz, P.; Fleischmann, O.; Vogel, A.; Hartmann, M.; Kromminga, H.; Bovensmann, H.; et al. Measurements of molecular absorption spectra with the SCIAMACHY pre-flight model: Instrument characterization and reference data for atmospheric remote-sensing in the 230–2380 nm region. *J. Photochem. Photobiol. A Chem.* **2003**, *157*, 167–184. [CrossRef]
42. Vandaele, A.; Hermans, C.; Simon, P.; Carleer, M.; Colin, R.; Fally, S.; Mérienne, M.; Jenouvrier, A.; Coquart, B. Measurements of the $NO_2$ absorption cross-section from 42,000 $cm^{-1}$ to 10,000 $cm^{-1}$ (238–1000 nm) at 220 K and 294 K. *J. Quant. Spectrosc. Radiat. Transf.* **1998**, *59*, 171–184. [CrossRef]
43. Wilmouth, D.M.; Hanisco, T.F.; Donahue, N.M.; Anderson, J.G. Fourier Transform Ultraviolet Spectroscopy of the A $2\Pi 3/2 \leftarrow$ X $2\Pi 3/2$ Transition of BrO. *J. Phys. Chem. A* **1999**, *103*, 8935–8945. [CrossRef]
44. Gierczak, T.; Burkholder, J.B.; Bauerle, S.; Ravishankara, A. Photochemistry of acetone under tropospheric conditions. *Chem. Phys.* **1998**, *231*, 229–244. [CrossRef]
45. Rufus, J.; Stark, G.; Smith, P.L.; Pickering, J.C.; Thorne, A.P. High-resolution photoabsorption cross section measurements of $SO_2$, 2: 220 to 325 nm at 295 K. *J. Geophys. Res. Planets* **2003**, *108*. [CrossRef]
46. Meller, R.; Moortgat, G.K. Temperature dependence of the absorption cross sections of formaldehyde between 223 and 323 K in the wavelength range 225–375 nm. *J. Geophys. Res. Atmos.* **2000**, *105*, 7089–7101. [CrossRef]
47. Barbero, A.; Blouzon, C.; Savarino, J.; Caillon, N.; Dommergue, A.; Grilli, R. A compact incoherent broadband cavity-enhanced absorption spectrometer for trace detection of nitrogen oxides, iodine oxide and glyoxal at levels below parts per billion for field applications. *Atmos. Meas. Tech.* **2020**, *13*, 4317–4331. [CrossRef]
48. Thalman, R.; Zarzana, K.J.; Tolbert, M.A.; Volkamer, R. Rayleigh scattering cross-section measurements of nitrogen, argon, oxygen and air. *J. Quant. Spectrosc. Radiat. Transf.* **2014**, *147*, 171–177. [CrossRef]
49. Fiedler, S.E.; Hese, A.; Ruth, A.A. Incoherent broad-band cavity-enhanced absorption spectroscopy. *Chem. Phys. Lett.* **2003**, *371*, 284–294. [CrossRef]
50. Platt, U.; Stutz, J. *Differential Optical Absorption Spectroscopy (DOAS)—Principles and Applications*; Springer: Berlin/Heidelberg, Germany, 2008; Volume 15. [CrossRef]
51. Dankaert, T.; Fayt, C.; Roozendael, M.V.; Smedt, I.D.; Letocart, V.; Merlaud, A.; Pinardi, G. QDOAS Software User Manual, Version 3.2. 2017. Available online: http://uv-vis.aeronomie.be/software/QDOAS (accessed on 31 January 2021).

52. Sander, S.P.; Friedl, R.R. Kinetics and product studies of the reaction chlorine monoxide + bromine monoxide using flash photolysis-ultraviolet absorption. *J. Phys. Chem.* **1989**, *93*, 4764–4771. [CrossRef]
53. Coburn, S.; Ortega, I.; Thalman, R.; Blomquist, B.; Fairall, C.W.; Volkamer, R. Measurements of diurnal variations and eddy covariance (EC) fluxes of glyoxal in the tropical marine boundary layer: Description of the Fast LED-CE-DOAS instrument. *Atmos. Meas. Tech.* **2014**, *7*, 3579–3595. [CrossRef]
54. Suhail, K.; George, M.; Chandran, S.; Varma, R.; Venables, D.; Wang, M.; Chen, J. Open path incoherent broadband cavity-enhanced measurements of $NO_3$ radical and aerosol extinction in the North China Plain. *Spectrochim. Acta Part A Mol. Biomol. Spectrosc.* **2019**, *208*, 24–31. [CrossRef]